\documentclass{iopjournal}
\usepackage{amsmath,amssymb,mathtools,siunitx,physics}

\newcommand{\prl}{Phys.~Rev.~Lett.}
\newcommand{\prc}{Phys.~Rev.~C}
\newcommand{\prd}{Phys.~Rev.~D}
\newcommand{\apj}{Astrophys.~J.}
\newcommand{\apjs}{Astrophys.~J.~Suppl.}

\newcommand{\aap}{Astron.~Astrophys.}

\begin{document}

\articletype{Paper} 

\title{Limiting cases of second-order moments of relativistic stars and their universality}

\author{Koutarou Kyutoku$^{1,2}$\orcid{0000-0003-3179-5216}}

\affil{$^1$Department of Physics, Graduate School of Science, Chiba University, Chiba 263-8522, Japan}

\affil{$^2$Interdisciplinary Theoretical and Mathematical Sciences Program (iTHEMS), RIKEN, Wako, Saitama 351-0198, Japan}

\email{kyutoku@chiba-u.jp}


\begin{abstract}
 Extending the work presented in a workshop ``From Quarks to Neutron Stars: Insights from kHz gravitational waves'', we discuss some limiting cases of the moment of inertia, tidal deformability, and spin-induced quadrupole moment for relativistic stars. First, conjecturing that a hierarchy of the length scale is the key to proposed universality among these second-order moments, we revisit the relation for incompressible relativistic stars (known as Schwarzschild's interior solution) as a candidate of the possible stiff limit. Second, we present the limiting form for the weak-field limit. In particular, we demonstrate how relativistic computations of tidal deformation are related to the traditional Newtonian counterpart, which might not have been presented explicitly in the literature.
\end{abstract}

\section{Introduction}

Second-order moments of relativistic stars such as neutron stars are important targets of gravitational-wave astronomy. Tidal deformability $\Lambda$ enables us to infer the properties of supranuclear-density matter and was actually measured from GW170817 \cite{2018PhRvL.121p1101A}. Because the tidal deformation becomes prominent at high frequency, future detectors with high sensitivity in the kilohertz range will become a powerful tool to improve our knowledge in this direction. In principle, gravitational waves also tell us about the spin-induced quadrupole moment of neutron stars. Currently, however, this quantity is usually expressed as a function of tidal deformability in gravitational-wave data analysis, even if the equation of state is sampled \cite{2018PhRvL.121p1101A}. This is justified by the so-called universal relation among the moment of inertia, tidal deformability, and spin-induced quadrupole moment \cite{2013Sci...341..365Y,2013PhRvD..88b3009Y} as long as the measurement error is larger than the systematic error of the universal relation, $\sim \order{1\%}$. Still, it is desirable to have an option of estimating the spin-induced quadrupole moment independently of the tidal deformability, particularly in future high-precision observations aided with space-borne gravitational wave detectors such as DECIGO \cite{2018PTEP.2018g3E01I}.

For the purpose of accelerating analysis that samples equations of state, we have presented a simplified reformulation for computing the spin-induced quadrupole moment \cite{2025PhRvD.112b4059K}. Using this reformulation, we obtain circumstantial evidence that the universality is approached when the quantities governing these second-order moments vary slowly inside the star (see, e.g., Refs.~\cite{2015ApJ...798..121S,2015PhRvD..91d4017C,2025PhRvD.112b3030K} for related arguments). Motivated by notions of universality encountered in other branches of physics, we further speculate that the universal relation may be a manifestation of hierarchy in length scales. That is, the universality is achieved when the variation length scale of physical quantities inside the star is much longer than the stellar radius. These discussions are presented in a workshop ``From Quarks to Neutron Stars: Insights from kHz gravitational waves.''

In this article, we extend our discussion in the workshop based on Ref.~\cite{2025PhRvD.112b4059K} to a specific investigation of a few limiting cases. In Sec.~\ref{sec:formulation}, we briefly review our reformulation for computing second-order moments of relativistic stars presented in Ref.~\cite{2025PhRvD.112b4059K}. As novel applications of this reformulation, Sec.~\ref{sec:stiff} and Sec.~\ref{sec:weak} discuss Schwarzschild's interior solution as a candidate of the possible stiff limit and the weak-field limit, respectively. In the latter, we explain in some detail how the Newtonian limit of relativistic computations of tidal deformability and the Love number presented in Ref.~\cite{2008ApJ...677.1216H} is related to the traditional computation of the Love number in Newtonian gravity, as it might not have been presented explicitly in the literature. Section \ref{sec:summary} is devoted to a summary. Throughout this paper, we adopt the geometric unit in which $G=c=1$. The energy density and pressure are denoted by $\varepsilon$ and $P$, respectively. The mass and radius of a star are denoted by $M$ and $R$, respectively, and the compactness is defined by $C \coloneqq M/R$. The quantities with a subscript $c$ denote the values at the stellar center.

\section{Reformulation} \label{sec:formulation}

We summarize the equations to derive second-order moments of relativistic stars based on our simplified reformulation \cite{2025PhRvD.112b4059K}. The background spherical configuration is derived by solving Tolman-Oppenheimer-Volkoff's equation (see, e.g., Sec.~6.2 of Ref.~\cite{1984ucp..book.....W}),
\begin{equation}
 \dv{P}{r} = - \frac{(\varepsilon + P) (m+4\pi P r^3)}{r(r-2m)} ,
\end{equation}
supplemented with the definition of (gravitational) mass
\begin{equation}
 \dv{m}{r} = 4\pi \varepsilon r^2 .
\end{equation}
We always assume that the equation of state $\varepsilon (P)$ is specified. The background metric is assumed to have the form
\begin{equation}
 \dd{s}^2 = -e^{2\nu} \dd{t}^2 + e^{2\lambda} \dd{r}^2 + r^2 (\dd{\theta}^2 + \sin^2 \theta \dd{\varphi}^2) ,
\end{equation}
and we readily obtain $e^{2\lambda} = (1-2m/r)^{-1}$. Tolman-Oppenheimer-Volkoff's equation is solved from the stellar center with $P(0) = P_c$ and $m(r \to 0) = (4\pi /3) \varepsilon_c r^3$. The integration is terminated at the stellar surface defined by the condition $P=0$. Once these equations are solved, the compactness $C$ allows us to write
\begin{equation}
 \nu = \frac{1}{2} \ln (1 - 2C) - H ,
\end{equation}
where the logarithm of specific enthalpy is defined by
\begin{equation}
 H = \int_0^P \frac{\dd{P'}}{\varepsilon (P') + P'} .
\end{equation}

The moment of inertia and spin-induced quadrupole moment are derived by solving equations for rotational perturbation \cite{1967ApJ...150.1005H,1968ApJ...153..807H}. The former is governed by a variable $w$, which obeys \cite{1987A&A...172...95Z}
\begin{equation}
 \dv{w}{r} = - \frac{w(w+3)}{r} + \frac{4\pi (w+4) (\varepsilon + P) r^2}{r-2m} .
\end{equation}
This is solved from the center with the regularity condition $w(r \to 0) = (16\pi /5) (\varepsilon_c + P_c) r^2$ to the surface to determine $w_s = w(R)$. The normalized moment of inertia is given by
\begin{equation}
 \bar{I} = \frac{w_s}{C^3 (2w_s + 6)} .
\end{equation}
Once the profile of $w$ is obtained (technically, we can solve for them all at once as far as the background has already been obtained), the spin-induced quadrupole moment can be computed by solving the system of two ordinary differential equations for variables $f$ and $\tilde{q}$ \cite{2025PhRvD.112b4059K},
\begin{align}
 \dv{f}{r} & = - \frac{2}{m+4\pi P r^3} f^2 + \Bqty{2 \dv{\nu}{r} - \frac{r^2}{m+4\pi P r^3} \bqty{4\pi (\varepsilon + P) - \frac{2m}{r^3}}} f + 2 \dv{\nu}{r} , \label{eq:dfdr} \\
 \dv{\tilde{q}}{r} & = - \bqty{\frac{2f}{m + 4\pi Pr^3} + \frac{2(w+1)}{r}} \tilde{q} + \frac{j^2 w^2}{6} \Bqty{\dv{\nu}{r} + \frac{1}{r} + \bqty{\dv{\nu}{r} - \frac{1}{2(m + 4\pi Pr^3)}} f} \notag \\
 & + \frac{8\pi (\varepsilon + P) r^3 j^2}{3(r-2m)} \Bqty{\dv{\nu}{r} + \frac{1}{r} + \bqty{\dv{\nu}{r} + \frac{1}{2(m+4\pi Pr^3)}} f} ,
\end{align}
where $j \coloneqq e^{-(\nu + \lambda)}$. The regularity condition requires $f(r \to 0) = (2\pi /3) (\varepsilon_c + 3P_c) r^2$ and $\tilde{q} (r \to 0) = (2\pi /3) (\varepsilon_c + P_c) e^{-2\nu_c} r^2$ simultaneously. Once $f_s = f(R)$ and $\tilde{q}_s = \tilde{q} (R)$ are obtained, the normalized spin-induced quadrupole moment is given by
\begin{equation}
 \bar{Q} = 1 + \frac{8}{5} \bqty{\frac{2Q_2^1 (\zeta_s)}{\sqrt{\zeta_s^2 - 1}} + f_s Q_2^2 (\zeta_s)}^{-1} \Bqty{\frac{(1-2C^3 \bar{I})^2}{C^2 \bar{I}^2} \tilde{q}_s + C^4 \bqty{1 - \pqty{1 + \frac{1}{C}} f_s}} ,
\end{equation}
where $\zeta_s \coloneqq 1/C - 1$ and $Q_l^m$ is the associated Legendre function of the second kind.

The tidal deformability is derived independently of these rotational perturbations. Specifically, we solve for tidal perturbation in the Regge-Wheeler gauge \cite{2008ApJ...677.1216H,2009PhRvD..80h4035D,2010PhRvD..82b4016P},
\begin{equation}
 \dv{y}{r} = - \frac{y^2}{r} + \frac{e^{2\lambda}}{r} [-1 + 4\pi (\varepsilon - P) r^2] y + r \bqty{4 \pqty{\dv{\nu}{r}}^2 + \frac{6e^{2\lambda}}{r^2} - 4\pi e^{2\lambda} \pqty{5\varepsilon + 9P + \frac{\varepsilon + P}{\dv*{P}{\varepsilon}}}} ,
\end{equation}
with the regularity condition $y(0) = 2$. The value of $y$ at the surface, $y_s = y(R)$, determines the dimensionless tidal deformability via
\begin{align}
 \Lambda & = \frac{16}{15} (1-2C)^2 [2 - y_s + 2C(y_s - 1)] \notag \\
 & \times \{2C [6 - 3y_s + 3C(5y_s - 8)] + 4C^3 [13 - 11y_s + C(3y_s - 2) + 2C^2 (1+y_s)] \notag \\
 & + 3 (1-2C)^2 [2 - y_s + 2C(y_s - 1)] \ln (1-2C)\}^{-1} .
\end{align}
It should be cautioned that, because of the jump in energy density at the stellar surface, $y_s = y(r \to R^-) - 3$ must be employed for incompressible stars \cite{2009PhRvD..80h4035D}.

\section{Stiff limit: Schwarzschild's interior solution} \label{sec:stiff}

In our previous work \cite{2025PhRvD.112b4059K}, we conjectured that the universality is achieved when the variation length scale of various quantities defined inside the star is much longer than the stellar radius. Although this speculation comes from numerical experiments for various polytropic indices, this may be compatible with the standard notion of ``universality'' encountered in other branches of physics, such as the critical phenomena and scattering problems. From this perspective, the universal relation for relativistic stars may be \emph{defined} by the stiffest possible matter.

\begin{table}
 \centering
 \caption{Coefficient of the fit, Eq.~\eqref{eq:fitI} and \eqref{eq:fitQ}, for Schwarzschild's interior solution.}
 \begin{tabular}{ccc}
  \hline
  $n$ & $c_I^n$ & $c_Q^n$ \\
  \hline \hline
  0 & \num{1.49579} & \num{-1.64137} \\
  1 & \num{0.0719351} & \num{0.705157} \\
  2 & \num{0.0161500} & \num{-0.0369580} \\
  3 & \num{6.55249e-4} & \num{9.56909e-4} \\
  4 & \num{-1.49925e-4} & \num{6.57490e-5} \\
  5 & \num{1.07434e-5} & \num{-7.98158e-6} \\
  6 & \num{-4.27147e-7} & \num{3.79668e-7} \\
  7 & \num{9.34875e-9} & \num{-9.17365e-9} \\
  8 & \num{-8.77189e-11} & \num{9.16662e-11} \\
  \hline
 \end{tabular}
 \label{table:fit}
\end{table}

One candidate for the stiffest possible matter is the incompressible fluid with $\varepsilon = \text{const}$. The incompressible stellar configuration in general relativity is known as Schwarzschild's interior solution \cite{1984ucp..book.....W},
\begin{equation}
 P(r) = \varepsilon \frac{\sqrt{1 - 2Mr^2/R^3} -\sqrt{1 - 2M/R}}{3 \sqrt{1 - 2M/R} - \sqrt{1 - 2Mr^2/R^3}} .
\end{equation}
From this viewpoint, we numerically derive $\bar{I}(\Lambda)$ and $\bar{Q}(\Lambda)$ for this solution and fit the results with eighth-degree polynomials by
\begin{align}
 \ln \bar{I} & = \sum_{n=0}^8 c_I^n (\ln \Lambda)^n , \label{eq:fitI} \\
 \ln (\bar{Q} - 1) & = \sum_{n=0}^8 c_Q^n (\ln \Lambda)^n , \label{eq:fitQ}
\end{align}
hoping that they will serve as a useful reference point [see Ref.~\cite{2015PhRvD..91d4017C} for analytic results of $\bar{I} (\Lambda)$]. The coefficients are listed in Table \ref{table:fit}. The order of polynomials is chosen to ensure that the average deviation is kept small for a wide range of $0.01 < C < 0.44$. Subtraction of the black-hole limit, $\bar{Q}_\mathrm{BH} = 1$, from $\bar{Q}$ happens to improve the quality of the fit, while that of $\bar{I}_\mathrm{BH} = 4$ from $\bar{I}$ does not help in our case. An advantage of this fit is that there is no arbitrariness in choosing the candidate equations of state except for our preference in favor of incompressibility. Obviously, these fitting formulae are independent from future updates of observational constraints or from proposals for new models.

\begin{figure}
 \centering
 \begin{tabular}{cc}
  \includegraphics[width=0.48\textwidth]{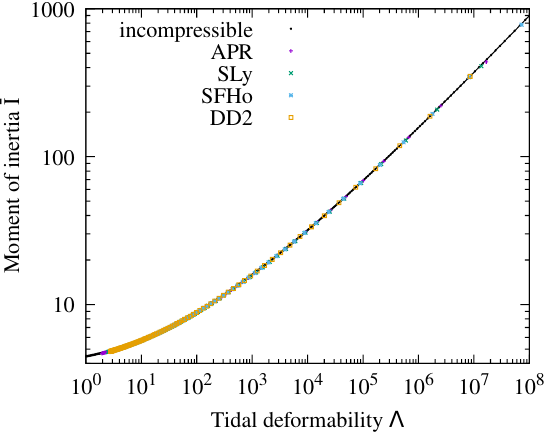} &
      \includegraphics[width=0.48\textwidth]{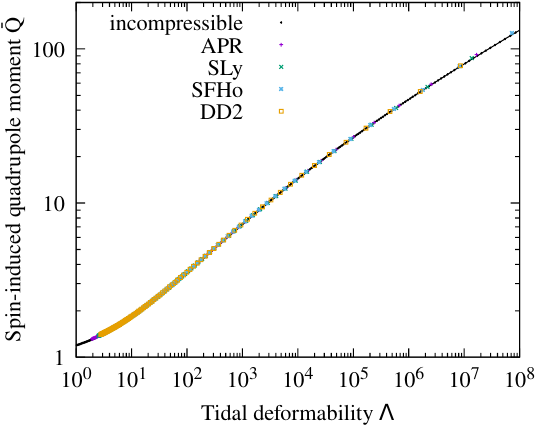}
 \end{tabular}
 \caption{Normalized moment of inertia, $\bar{I}$ (left), and spin-induced quadrupole moment, $\bar{Q}$ (right), as a function of dimensionless tidal deformability, $\Lambda$. We show both data points for Schwarzschild's interior solution (black dot, denoted by ``incompressible'') and the fit given by Eqs.~\eqref{eq:fitI} and \eqref{eq:fitQ} (black solid curve). They agree nearly perfectly on the level of these plots. We adopt APR \cite{1998PhRvC..58.1804A}, SLy \cite{2001A&A...380..151D}, SFHo \cite{2013ApJ...774...17S}, and DD2 \cite{2014ApJS..214...22B} as the models for selected nuclear-theory-based equations of state.} \label{fig:data}
\end{figure}

The results for incompressibile stars approximate second-order moments of relativistic stars computed by employing selected nuclear-matter equations of state with reasonable precision. Figure~\ref{fig:data} presents $\bar{I}(\Lambda)$ and $\bar{Q}(\Lambda)$ for a wide range of dimensionless tidal deformability, $1 \le \Lambda \le 10^8$. The agreement among various models is remarkable as demonstrated by many previous studies \cite{2013Sci...341..365Y,2013PhRvD..88b3009Y}.

\begin{figure}
 \centering
 \begin{tabular}{cc}
  \includegraphics[width=0.48\textwidth]{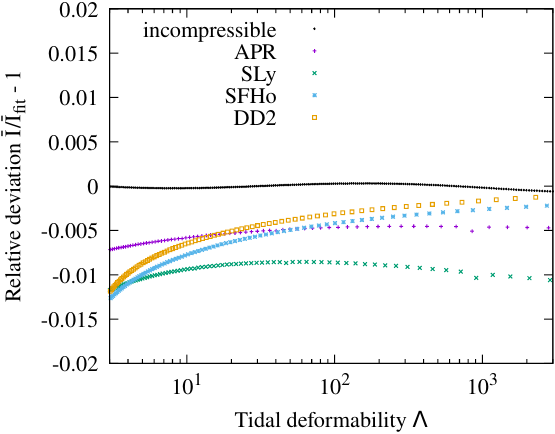} &
      \includegraphics[width=0.48\textwidth]{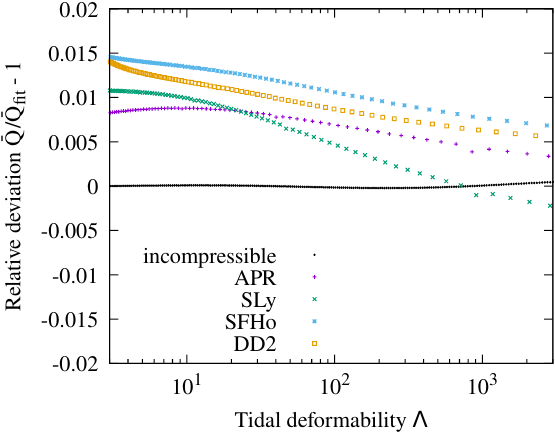}
 \end{tabular}
 \caption{Same as Fig.~\ref{fig:data} but shown as relative deviations from the fit, Eqs.~\eqref{eq:fitI} and \eqref{eq:fitQ}, in the restricted range of dimensionless tidal deformability. The maximum relative deviations in this range of $\bar{I}(\Lambda)$ are $0.8\%$, $1.2\%$, $1.3\%$, and $1.2\%$ for APR4, SLy, SFHo, and DD2, respectively, and those of $\bar{Q}(\Lambda)$ are $0.9\%$, $1.1\%$, $1.5\%$, and $1.4\%$, respectively.} \label{fig:diff}
\end{figure}

A close inspection reveals that the incompressible results deviate systematically from those for nuclear-theory-based equations of state. Figure~\ref{fig:diff} shows the deviation of numerical results from the incompressible fits, Eqs.~\eqref{eq:fitI} and \eqref{eq:fitQ}, for $3 \le \Lambda \le \num{3000}$, which may be representative of realistic neutron stars. The fit reproduces numerical results for Schwarzschild's interior solution within $\lesssim 0.1\%$ in this range, indicating the degree of accuracy of this fit. Numerical results for nuclear-theory-based equations of state deviate by a larger amount of $\sim 1\%$ from this fit. This degree of deviation is consistent with universal relations proposed by fitting various stellar models \cite{2013Sci...341..365Y,2013PhRvD..88b3009Y}. In addition, we find that nuclear-theory-based results are systematically smaller for $\bar{I}$ and larger for $\bar{Q}$ than incompressible results. If the incompressibility serves as an appropriate definition of universality, this systematic deviation is interpreted as the failure of relativistic stars governed by nuclear-theory-based equations of state to achieve exact universality rather than mere equation-of-state dependence.

We caution that, even if we accept the idea that the stiffest possible matter defines the universality, the incompressible matter might not be the suitable candidate. As is well known, incompressibility implies the infinite speed of sound, which violates causality. Accordingly, Schwarzschild's interior solution might be ruled out as a viable candidate of relativistic stars even in our context. Other candidates may include matter with luminal sound propagation, $P = \varepsilon + \text{const}$. Because the luminal equation of state lies between the incompressible and nuclear-theory-based models, relative deviations is expected to be smaller than those presented in Fig.~\ref{fig:diff}, i.e., less than $1\%$ from both models, unless the constant is chosen to be an extraordinay value. In this study, however, we refrain from trying out various possible candidates. If our view on universality is reasonably valid, in-depth analysis of relevant differential equations, e.g., by renormalization group theory, could give us an idea of matter at the critical point.

\section{Weak-field limit: (post-)Newtonian gravity} \label{sec:weak}

The weak-field, Newtonian limit is not only useful on its own but also gives us an intuitive understanding of complicated relativistic results. In this section, we recover $G$ and $c$ for clarity. In the Newtonian limit, the energy density $\varepsilon$ reduces to $\rho c^2$ with $\rho$ being the rest-mass density. Tolman-Oppenheimer-Volkoff's equation reduces to the usual equation for the hydrostatic equilibrium,
\begin{equation}
 \dv{P}{r} \approx - \frac{G \rho m}{r^2} \label{eq:dpdrN}
\end{equation}
with $\dv*{m}{r} \approx 4\pi \rho r^2$. It is worthwhile to note that $c^2 \nu$ is identical to the gravitational potential $\Phi$ in this limit and satisfies the Newtonian Bernoulli's theorem, $\Phi + \int \dd{P}/\rho \approx - GM/R$.

The weak-field limit of $\dv*{w}{r}$ is
\begin{equation}
 \dv{w}{r} \approx - \frac{3w}{r} + \frac{16\pi G\rho r}{c^2}
\end{equation}
with
\begin{equation}
 \bar{I} \approx \frac{w_s}{6C^3} .
\end{equation}
Because $w = \order{c^{-2}}$, this may be regarded as an equation at the first post-Newtonian order for gravitomagnetic fields. The moment of inertia is of course defined also in Newtonian gravity, and this apparent relativistic nature comes from the fact that it is defined from the angular momentum of the spacetime in general relativity. An intuitive expression in terms of an integration over the fluid is also provided in Ref.~\cite{1967ApJ...150.1005H}.

The equations for the spin-induced quadrupole moment reduce to
\begin{align}
 \dv{f}{r} & \approx - \frac{2c^2}{Gm} f^2 + \pqty{\frac{2}{r} - \frac{4\pi \rho r^2}{m}} f + \frac{2Gm}{c^2 r^2} , \\
 \dv{\tilde{q}}{r} & \approx - \frac{2\tilde{q}}{r} \pqty{1 + \frac{c^2 f r}{Gm}} + \frac{8\pi G\rho r}{3c^2} \pqty{1 + \frac{c^2 fr}{2Gm}} .
\end{align}
These equations are also $\order{c^{-2}}$. It is worthwhile to note that, once we replace $f$ by $\eta \coloneqq 2c^2 fr/(Gm) - 1$, the former equation turns out to be identical with Clairaut-Radau's equation of degree $l=2$ (see, e.g., Sec.~2.4 of Ref.~\cite{2014grav.book.....P}),
\begin{equation}
 r \dv{\eta}{r} + \eta (\eta - 1) + \frac{6\rho}{\bar{\rho}} (\eta + 1) - l(l+1) = 0 , \label{eq:radau}
\end{equation}
where $\bar{\rho} \coloneqq 3m/(4\pi r^3)$. This suggests that Eq.~\eqref{eq:dfdr} may be rewritten more concisely by appropriately changing the variable, and that even tidal deformability might be derived from this equation (see below), but the author has not yet attempted to do so. The value of the normalized quadrupole moment is given by
\begin{equation}
 \bar{Q} \approx \frac{2\tilde{q}_s}{C^5 (C+2f_s) \bar{I}^2} .
\end{equation}

The Newtonian limit of the tidal perturbation is given by \cite{2008ApJ...677.1216H,2009ApJ...697..964H}
\begin{equation}
 \dv{y}{r} = - \frac{y(y+1)}{r} + \frac{6}{r} - \frac{4\pi G\rho r}{\dv*{P}{\rho}} \label{eq:dydrN}
\end{equation}
and
\begin{equation}
 \Lambda = \frac{2-y_s}{3(y_s+3) C^5} .
\end{equation}
We also show explicitly that this Newtonian limit is equivalent to the traditional Newtonian perturbation in terms of Clairaut-Radau's equation. If we introduce a new variable,
\begin{equation}
 \eta = y + 1 - \frac{4\pi \rho r^3}{m} , \label{eq:transf}
\end{equation}
it is readily found that Eq.~\eqref{eq:dydrN} is equivalent to Eq.~\eqref{eq:radau} for $\eta$ defined here and that the dimensionless tidal deformability is given in terms of $\eta_s = \eta (R)$ by
\begin{equation}
 \Lambda = \frac{3-\eta_s}{3(\eta_s + 2) C^5} .
\end{equation}
This equation agrees with the classical definition of tidal Love number, $k_l = (3-\eta_s)/[2(\eta_s + l)]$, for $l=2$ \cite{2014grav.book.....P}.

The origin of Eq.~\eqref{eq:transf} may be better understood by going back to the second-order differential equation for the metric perturbation $H_0$, which is defined by $\delta g_{tt} \coloneqq - c^2 e^{2\nu} H_0$ and indeed defines $y_l$ via $y_l = \dv*{\ln H_{0,lm}}{\ln r}$ under the spherical harmonic expansion $H_0 = \sum_{lm} H_{0,lm} Y^{lm}$. Here, we generalize the discussion to all the modes with $l \ge 2$ to make the situation transparent. In the Newtonian limit, this metric perturbation obeys \cite{2008ApJ...677.1216H}
\begin{equation}
 \dv[2]{H_{0,lm}}{r} + \frac{2}{r} \dv{H_{0,lm}}{r} + \bqty{\frac{4\pi G\rho}{\dv*{P}{\rho}} - \frac{l(l+1)}{r^2}} H_{0,lm} = 0 . \label{eq:d2hdr2}
\end{equation}
The counterpart in Newtonian gravity of this equation is understood by recalling that $c^2 H_0 /2$ in the Newtonian limit represents the perturbation of the gravitational potential, $\delta \Phi = \sum_{lm} \Phi_{lm} Y^{lm}$. By expanding the density perturbation as $\delta \rho = \sum_{lm} \rho_{lm} Y^{lm}$, perturbed Poisson's equation derives
\begin{equation}
 \dv[2]{\Phi_{lm}}{r} + \frac{2}{r} \dv{\Phi_{lm}}{r} - \frac{l(l+1)}{r^2} \Phi_{lm} = 4\pi G \rho_{lm} .
\end{equation}
Because the Bernoulli's theorem, $\Phi + \int \dd{P}/\rho = \text{const.}$, holds even under the presence of static tidal fields, we have $\rho_{lm} = -[\rho / (\dv*{P}{\rho})] \Phi_{lm}$ for $l \neq 0$ modes. By eliminating $\rho_{lm}$ in favor of $\Phi_{lm}$, the perturbed potential is shown to obey the equation
\begin{equation}
 \dv[2]{\Phi_{lm}}{r} + \frac{2}{r} \dv{\Phi_{lm}}{r} + \bqty{\frac{4\pi G\rho}{\dv*{P}{\rho}} - \frac{l(l+1)}{r^2}} \Phi_{lm} = 0 , \label{eq:d2Phidr2}
\end{equation}
which exactly reproduces Eq.~\eqref{eq:d2hdr2}.

Clairaut's equation governs the fractional radial displacement of a constant density surface, $\delta r = r \sum_{lm} \xi_{lm} Y^{lm}$. Here, $\xi_{lm}$ is taken to be dimensionless, although this symbol is conventionally used to represent the radial displacement itself (see, e.g., Ref.~\cite{2014grav.book.....P}). It follows from this definition that the Eulerian variation of the rest-mass density is given by $\rho_{lm} = - r \xi_{lm} \dv*{\rho}{r}$. Thus, the potential perturbation is expressed in terms of $\xi_{lm}$ as
\begin{equation}
 \Phi_{lm} = - \frac{Gm}{r} \xi_{lm} ,
\end{equation}
with the aid of Eq.~\eqref{eq:dpdrN}. Substitution of this equation into Eq.~\eqref{eq:d2Phidr2} leads to Clairaut's equation \cite{2014grav.book.....P},
\begin{equation}
 r^2 \dv[2]{\xi_{lm}}{r} + \frac{6\rho}{\bar{\rho}} \pqty{r \dv{\xi_{lm}}{r} + \xi_{lm}} - l(l+1) \xi_{lm} = 0 .
\end{equation}
This is equivalent to Clairaut-Radau's equation, Eq.~\eqref{eq:radau}, for $\eta_l = \dv*{\ln \xi_{lm}}{\ln r}$. By expressing $\dv*{\ln \xi_{lm}}{\ln r}$ in terms of $\Phi_{lm}$ that is identified with $H_{0,lm}$ up to a constant factor, we arrive at Eq.~\eqref{eq:transf}.

Finally, we comment on the results for incompressible Newtonian stars. Remarkably, the equations for $w$, $f$, $\tilde{q}$, and $y$ are all satisfied by the leading-order behavior at the stellar center. The surface values are readily found to be $w_s = 12C/5$, $f_s = \tilde{q}_s = C/2$, and $y_s = -1$ after the surface correction. The values of second-order moments are $\bar{I} = 2/(5C^2)$, $\bar{Q} = 25/(8C)$, and $\Lambda = 1/(2C^5)$. If we presume that they represent the Newtonian limit of universality, we have $\bar{I} = (2^{7/5}/5) \Lambda^{2/5}$ and $\bar{Q} = (25/2^{14/5}) \Lambda^{1/5}$ \cite{2013PhRvD..88b3009Y}.

\section{Summary} \label{sec:summary}

We discussed a few limiting cases of second-order moments of relativistic stars and the proposed universality among them. Our reformulation \cite{2025PhRvD.112b4059K} simplifies the numerical computation and hopefully serves as a step toward clear understanding of the relativistic stellar perturbation. We speculate that the origin of universality may be related to the stiffness of matter, and we presented the result for incompressible, Schwarzschild's interior solution as a candidate of possible stiff limit. We also took this opportunity to discuss the weak-field limit, focusing particularly on the relation between relativistic tidal perturbations and the traditional Newtonian computation based on Clairaut-Radau's equation.

\ack{We thank Hayato Miyazono and Kent Yagi for valuable discussions.}

\funding{This work is supported by JSPS KAKENHI Grant-in-Aid for Scientific Research No.~JP26K07062.}


\bibliographystyle{iopart-num}
\providecommand{\newblock}{}

\end{document}